# Flyby Anomaly in the Variation Principle of General Relativity

Shubhen BISWAS[ɫ]

The anomalous velocity deviation in the osculating planetary flyby attracts enough attention as a problem of General Relativity. In connection of rotating weak field massive source the Lense-Thirring metric is diagonalized to find the equation of motion from action invariance Hamilton principle in pure relativistic theory. The computation for near Earth flyby shows energy anomaly over the asymptotic in and out velocity obeying Anderson's empirical formula!

[ɫ] Université de Tours, Parc de Grandmont, 37200 Tours, France.
  Email: - shubhen3@gmail.com

# 1. Introduction:

The Voyager missions 1970 and1980 by the Jet Propulsion Laboratory (JPL) became the central attractions for gravity assist planetary flyby! Investigations by Antreasian, P.G. and J.R. Guinn, 1998, presented the unexpected delta-$v$ increases during the Earth gravity assists flybys, Galileo and NEAR [1]. Extensive work done over gravity assist flybys in data analysis have been done by Anderson et al. [2, 3]. The velocity deviation in planetary flyby is regarded as one of the several gravitational anomalies in the solar system [4]. If we synthesis the flyby anomalies it has certain features during flybys as described by Anderson et al. one of the key investigators [2, 3].

a) For an osculating spacecraft in its hyperbolic geocentric trajectory can both give kinetic energy to a spacecraft boosting orbital velocity and also can take kinetic energy that in lowering the velocity!

b) If a space probe grazing inside the planetary orbit, travels behind the planet counter-clockwise, kinetic energy will be increased. Contrarily if a satellite comes from inside the planetary orbit, travels in front clockwise, kinetic energy will be decreased.

c) on a seminal paper of Anderson et al. 2008 [3] have proposed the phenomenological empirical formula:

$$\left.\frac{\Delta V}{V}\right|_\infty = K(\cos \delta_i - \cos \delta_o) \qquad (1)$$

In this empirical formula the asymptotic velocity $V_\infty$ of its hyperbolic trajectory, $\delta_i$ and $\delta_o$ are the declinations for the incoming and outgoing osculating velocity vectors and K is a constant. The value of K seems to be close to $\frac{2\omega_\oplus R_\oplus}{c}$, where $\omega_\oplus$ is the angular velocity **Fig.1** of the Earth, $R_\oplus$ is the Earth radius and $c$ is the speed of light.

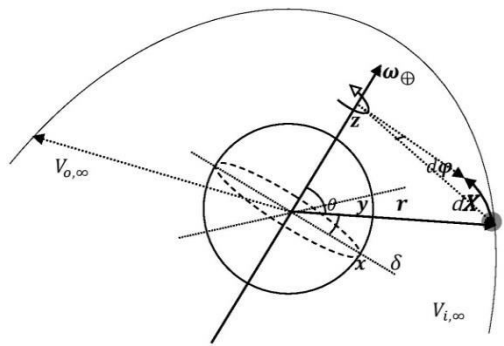

**Fig.1** Flyby trajectory near Earth

Analysing the Doppler radio-metric frequency tracking data for a number of Earth gravity assists spacecrafts confirms the unexpected energy change that occurs during the flyby. In case of the Juno space probe this anomalous behaviour during near Earth flyby was reported with increase in the asymptotic velocity $V_\infty$ of its hyperbolic trajectory was $\Delta V_\infty$ = 3.92±0.08 mm/s [5].

The paper by P.G. Antreasian and J.R. Guinn,1998, **[6]** speculated some non–standard physical models they are Non–conservative potential energy, likely the Non–Newtonian gravity **[7]**, post parametric Newtonian (PPN), Modifications of relativity and Torsion model but these are not compatible to the stability of the planetary orbit. The flyby anomaly may be attributed to earth-bound dark matter **[8]**. L. Acedo, 2018, **[9]** explored the idea of a strong gravitomagnetic field as the source of the flyby anomaly. This attracts huge interest regarding perturbation of highly eccentric orbital trajectory of the spacecrafts. But numerical model shows inconsistency in modelling. Also the first post-Newtonian (1PN) gravitomagnetic (GM), Lense-Thirring component of the Earth's field doesn't fit with data analysis softwares **[4]** on the Juno's. Since then, many alternatives have been considered and dismissed **[10]** and some with local and cosmological implications, such as torsion models, are still a field of active research **[11]**, **[12].**

The flyby anomaly is still a puzzle yet to be solved beyond the standard methods of modern theoretical astrodynamics on a pure phenomenological context. In the following sections we will be carving a new theoretical model for the dynamics of the flyby applying the general covariance principle of the General Relativity (GR) using invariance in the action principle. Let us first starting to sharpen our tools!

## 2. The space time metric for rotating massive body:

So far a solution for the Einsteins field equation brings beautifully the static isotropic metric as given by Schwarzschild **[13, 14, 15]**

$$ds^2{}_{Sch.} = -\left(1 - \frac{2GM}{c^2 r}\right) c^2 dt^2 + \left(1 - \frac{2GM}{c^2 r}\right)^{-1} dr^2 + r^2 d\theta^2 + r^2 \sin^2\theta \, d\varphi^2 \qquad (2)$$

Eq. (2) can be described excluding higher order perturbation in the PPN in quasi Minkowskian co-ordinates following transformations **[14, 15]**, $x^1 = r \sin\theta \cos\varphi$, $x^2 = r \sin\theta \sin\varphi$, $x^3 = r \cos\theta$, and $dX^2 = \sum_{i=1}^{3}(dx^i)^2$

$$ds^2{}_{Sch.} = -\left[1 - \frac{2GM}{c^2 r}\right] c^2 dt^2 + \left[1 + \frac{2GM}{c^2 r}\right] dX^2 \qquad (3)$$

It clearly got success in forecasting the light bending or the precision of perihelion. But to study most of the heavenly bodies like rotating black hole or planets we need the metric for a rotating axisymmetric source. The Kerr metric **[16]** will be a good choice in this matter.

In a convenient Boyer and Lindquist **[17]** coordinate the metric with only one off-diagonal terms:

$$ds^2{}_{Kerr} = -\left(1 - \frac{2GM}{c^2 \Sigma}\right) c^2 dt^2 - \frac{4GMj \, r \sin^2\theta}{c^2 \Sigma} dt d\varphi + \frac{\Sigma}{\Delta} dr^2 + \Sigma d\theta^2 +$$

$$\left(r^2 + a^2 + \frac{2GMj^2 \, r \sin^2\theta}{c^2 \Sigma}\right) \sin^2\theta \, d\varphi^2 \qquad (4)$$

$$\Sigma = r^2 + a^2 cos^2\theta \; ; \Delta = r^2 + a^2 - \frac{2GMr}{c^2} \; ; j = \text{angular momentum per unit mass}$$

Manifestly the Kerr metric in Eq.(4) has only two Killing fields: the time-like field makes it stationary and a field generating rotations about the axis of rotation keeps it axis symmetry. But it is not invariant under time reversal, $t \to -t$, hence is not static. Only at $r \to \infty$ the time Killing vector is hypersurface orthogonal and can be static.

The metric outside for a weak-field, slowly rotating source is approximately **[16]**

$$ds^2{}_{LT} = ds^2{}_{Sch.} - \frac{4GMj\, sin^2\theta}{c^2 r} dt d\varphi \quad (5)$$

In concern of weak-field objects rotating slowly as the Earth like planets the Lense-Thirring (LT) metric in Eq.(5) for "dragging of inertial frames" is most suitable to study orbital motion of the bound osculating flyby.

To ease the handling of metric elements **[18, 19, 20]** later in an algebraic matrix form here we will switch over from spherical polar to Cartesian coordinates to rewrite Eq.(5) in quasi Minkowskian representation with local temporal element, $dx^0 = cdt$

$$ds^2{}_{LT} = -\left[1 - \frac{2GM}{c^2 r}\right](dx^0)^2 + \left[1 + \frac{2GM}{c^2 r}\right]dX^2 - \frac{4GMj}{c^3 r^3}[x^1 dx^2 - x^2 dx^1]dx^0 \quad (6)$$

Let us define spatial coordinates in $x^1 = x; \; x^2 = y; \; x^1 = z$ and two parameters $b = \frac{2GM}{c^2 r}$ and $a = \frac{2GMj}{c^3 r^3}$

Now the LT metric Eq.(6) in matrix form turns into

$$[g_{\mu\nu}]_{LT} = \begin{bmatrix} -1+b & ay & -ax & 0 \\ ay & 1+b & 0 & 0 \\ -ax & 0 & 1+b & 0 \\ 0 & 0 & 0 & 1+b \end{bmatrix} \quad (7)$$

**Diagonalization of metric:**

In getting dynamics equation of motion for significantly low mass objects like spacecraft or planetary satellite under influence of massive rotating celestial body needs pure GR theory! The geodesic motion of the spacecrafts due to curved spacetime manifested by the massive planets requires Christoffel connections. But each connection element $\Gamma^\mu_{\rho\sigma}$ is associated with first order metric derivations, so it is difficult to deal with the above metric Eq.(7) having off diagonal elements!

$$\Gamma^\mu_{\rho\sigma} = \frac{1}{2} g^{\mu\nu}\left(\frac{\partial g_{\nu\sigma}}{\partial x^\rho} + \frac{\partial g_{\nu\rho}}{\partial x^\sigma} - \frac{\partial g_{\rho\sigma}}{\partial x^\nu}\right) \quad (8)$$

The need of diagonalizations have been practised in 2D case of Vaidya metric**[21]**also the diagonalization of metric have been a central crucial place to get easy solution in GR.

Following the Non Local Point Transformations NLPT **[22]** has been proposed for diagonalization of the Kerr metric.

If we consider at a point on the rotating body an equal and opposite angular velocity in backward time i.e. likely to just freezing the angular motion then we can avoid the off-diagonal cross terms in the spacetime metric. In such case in the same time to keep the world line element $ds$ invariant both temporal and spatial parts will get perturbed to compensate the stretching of the metric space.

Generalization of flat space-time metric transformations to curved space-time metric in geometric perturbation by Biswas[23] we can represent any curved spacetime metric $[g_{\mu\nu}]$ by quasi Lorentz transformations of a flat Minkowskian metric $[\eta_{\mu\nu}]$

$$[g_{\mu\nu}] = \Lambda [\eta_{\mu\nu}] \qquad (9)$$

$$[\eta_{\mu\nu}] = \begin{pmatrix} -1 & 0 & 0 & 0 \\ 0 & 1 & 0 & 0 \\ 0 & 0 & 1 & 0 \\ 0 & 0 & 0 & 1 \end{pmatrix} \qquad (10)$$

$\Lambda$ is a matrix connected through perturbation metric $[h_{\mu\nu}]$

$$\Lambda = \left[\mathbb{1} + [h_{\mu\nu}][\eta_{\mu\nu}]^{-1}\right] \qquad (11)$$

The algebra to compute of diagonalization of arbitrary matrix $[A]$ is given by the straight forward condition

$$[A]X = \lambda X \qquad (12)$$

In consideration of Eq.(12) above identity $[A]$ and $X$ having same dimension, we can express

$$\Lambda [\eta_{\mu\nu}] = \lambda [\eta_{\mu\nu}] \qquad (13)$$

$$[g_{\mu\nu}] = \lambda [\eta_{\mu\nu}] \qquad (14)$$

$$\det. [g_{\mu\nu} - \lambda \eta_{\mu\nu}] = 0 \qquad (15)$$

The Eigen value condition in the metric diagonalization in Eqns.(14) and (15) appears little difference with presence of matrix of flat spacetime metric $[\eta_{\mu\nu}]$ in place of the identity matrix$[\mathbb{1}]$ unlike generally used in elementary algebra!

Since the Minkowski metric $[\eta_{\mu\nu}]$ by construction describes a flat spacetime only having diagonal elements then an arbitrary spacetime metric $[g_{\mu\nu}]$ can be thought of equivalently as a diagonalized metric, just we need to find corresponding Eigen values!

Applying diagonalization condition Eq.(15) in Eq.(7) of $[g_{\mu\nu}]_{LT}$ leads

$$\begin{vmatrix} -1+b+\lambda & ay & -ax & 0 \\ ay & 1+b-\lambda & 0 & 0 \\ -ax & 0 & 1+b-\lambda & 0 \\ 0 & 0 & 0 & 1+b-\lambda \end{vmatrix} = 0 \quad (16)$$

The determinant of Eq.(16) implies

$$[(-1+b+\lambda)(1+b-\lambda) - a^2(y^2+x^2)](1+b-\lambda)(1+b-\lambda) = 0 \quad (17)$$

Taking, $a^2(y^2+x^2) = a^2 r^2 \sin^2\theta = A^2$ (18)

The first factor of Eq.(17) $[\lambda^2 - 2\lambda + 1 - b^2 + A^2] = 0$ (19)

$$\lambda = 1 \pm (b+A)\left[1 - \frac{2A}{b+A}\right]^{\frac{1}{2}} \quad (20)$$

Using back to the conditions in between Eqns.(6) and (7)

$$A = a\, r \sin\theta = \frac{2GMj}{c^3 r^2} \sin\theta \quad (21)$$

$GM_\oplus = 398600.4415\ km^3/s^2$; Earth radius, $R_\oplus = 6{,}378.1363$ km and mass taken as $M_\oplus = 5.9636 \times 10^{24}$ kg. [3] angular velocity $\omega_\oplus = \frac{2\pi}{86400}\ radian/sec$.

Taking Earth as a solid sphere of angular momentum $j_\oplus$ of

$$j_\oplus = \frac{2}{5}\omega_\oplus R_\oplus^{\,2} \quad (22)$$

$$\left(\frac{A}{b}\right)_\oplus = \frac{j}{cr}\sin\theta \sim \frac{4}{r}\sin\theta \quad (23)$$

Eqns. (22) and (23) implies, $A_\oplus \ll b_\oplus$ and ignoring the square bracketed term in Eq.(20) we have Eigen values for the time like LT metric

$$\lambda_0 = (-1+b+A) \quad (24)$$

$$\lambda_1 = (1+b+A) \quad (25)$$

$$\lambda_{2,3} = (1+b) \quad (26)$$

But without any loss of generality of the metric itself we can arrange its spatial Eigen values in any order, i.e. the spatial Eigen elements of the diagonal matrix are interchangeable! For each of the different arrangements we have simultaneous solutions. Finally diagonalization of $[g_{\mu\nu}]_{LT}$ reduces to three different equivalent forms

$$[g_{\mu\nu}]^{(1)}{}_{DLT} = \begin{bmatrix} -1+b+A & 0 & 0 & 0 \\ 0 & 1+b+A & 0 & 0 \\ 0 & 0 & 1+b & 0 \\ 0 & 0 & 0 & 1+b \end{bmatrix} \quad (27)$$

$$[g_{\mu\nu}]^{(2)}{}_{DLT} = \begin{bmatrix} -1+b+A & 0 & 0 & 0 \\ 0 & 1+b & 0 & 0 \\ 0 & 0 & 1+b+A & 0 \\ 0 & 0 & 0 & 1+b \end{bmatrix} \quad (28)$$

$$[g_{\mu\nu}]^{(3)}{}_{DLT} = \begin{bmatrix} -1+b+A & 0 & 0 & 0 \\ 0 & 1+b & 0 & 0 \\ 0 & 0 & 1+b & 0 \\ 0 & 0 & 0 & 1+b+A \end{bmatrix} \quad (29)$$

$$\det.[g_{\mu\nu}]_{LT} = g = -\lambda_0 \lambda_1 \lambda_2 \lambda_3 \quad (30)$$

The above diagonalization method keeps the four volume element, $\sqrt{|g|}dx^0 dx dy dz$ invariant for each representation of Eqns. (27), (28) and (29) as with parents one, is justified reasonably in GR. The metric proper time is same for all the three simultaneous solutions! In terms of new diagonalization proper time $d\tau$ related with coordinates time $dt$ as

$$d\tau^2 = (1-b-A)dt^2 \quad (31)$$

**The general relativistic action principle:**

The orbital trajectory of any object under the presence of matter energy of a heavenly body is described by the geodesic motion in general relativity. For geodesics as extremal paths based on the standard variation principle we can have over Hamilton's action invariance principle [13, 14].

Let us choosing a normalized Lagrangian[13]

$$\mathcal{L} = g_{\mu\nu} \dot{x}^\mu \dot{x}^\nu = \epsilon, \quad \dot{x}^\mu = \frac{dx^\mu}{d\tau} \quad (32)$$

And action, $$S = \int \left( g_{\mu\nu} \frac{dx^\mu}{d\tau} \frac{dx^\nu}{d\tau} \right) d\tau \equiv \int \mathcal{L} d\tau \quad (33)$$

This Eq.(33) on variation satisfies the Euler Lagrange equation and shows geodesic motion as free fall

$$\frac{d}{d\tau} \left[ \frac{\partial \mathcal{L}}{\partial (dx^\mu/d\tau)} \right] = \frac{\partial \mathcal{L}}{\partial x^\mu} \quad (34)$$

$$\frac{d^2 x^\rho}{d\tau^2} + \Gamma^\rho_{\mu\nu} \frac{dx^\mu}{d\tau} \frac{dx^\nu}{d\tau} = 0 \quad (35)$$

The proper time derivative of the Lagrangian gives

$$\frac{d[\mathcal{L}]}{d\tau} = \frac{1}{2}[v^\rho \partial_\rho (g_{\mu\nu}) + (\partial_\mu v^\rho) g_{\rho\nu} + (\partial_\nu v^\rho) g_{\rho\mu}] \frac{dx^\mu}{d\tau} \frac{dx^\nu}{d\tau} \quad (36)$$

$$\frac{d[\mathfrak{L}]}{d\tau} = \frac{1}{2}[L_v g]_{\mu\nu} \frac{dx^\mu}{d\tau} \frac{dx^\nu}{d\tau} \qquad (37)$$

Now if $v$ is a Killing vector field (diagonalization leaves metric static and Killing vector field, $\partial_t$) the Lie derivative of the metric vanishes $[L_v g]_{\mu\nu} = 0$ **[13]**, we have the conservation law for the generalised Lagrangian.

$$\frac{d[\mathfrak{L}]}{d\tau} = 0 \qquad (38)$$

Thus the Lagrangian is symmetric under evolution parameter $\tau$ and following Noethers theorem the symmetry requires a constant of motion, here it is the total energy or Hamiltonian $\mathcal{H}(p, x, \tau)$ in language of mechanics.

$$\mathcal{H}(p, x, \tau) = g^{\mu\nu} p_\mu p_\nu \qquad (39)$$

For non zero real mass objects like flyby spacecrafts $\mathcal{H} > 0$ stands for time like geodesic!

The Hamiltonian can be sought by a Legendre transformation of the Lagrangian,

$$\mathcal{H}(p, x, \tau) = p_\mu \frac{dx^\mu}{d\tau} - \mathfrak{L}(x, \dot{x}, \tau) \qquad (40)$$

The generalised covariant momentum $p_\mu$ and contravariant four velocity $\frac{dx^\mu}{d\tau}$, above Eq.(40) could lead us in getting equation of motion by taking the variation of the Hamiltonian between two different points along the geodesic, which is our intended trajectory of the osculating flyby!

$$\Delta\mathcal{H}(p, x, \tau) = \Delta\left[p_\mu \frac{dx^\mu}{d\tau}\right] - \Delta\mathfrak{L}(x, \dot{x}, \tau) \qquad (41)$$

Since the Lagrangian and conjugated Hamiltonian are both in energy unit and they are scalar invariant in any generalised coordinate systems along the geodesic as in Eq.(41)

$$\Delta\mathcal{H}(p, x, \tau) = \Delta\mathfrak{L}(x, \dot{x}, \tau) = 0 \qquad (42)$$

Eqns.(41) and (42) implies along geodesic, $\qquad \Delta\left[p_\mu \frac{dx^\mu}{d\tau}\right] = 0 \qquad (43)$

This Eq.(43) is just the general invariance of scalar product of four velocity.

**Computation of Flyby anomaly:**

Following **Fig.1** let us think for the flyby it has two asymptotes for in and out velocity one at the points $(r, \theta_i)$ and $(r, \theta_o)$ respectively. We assume only polar angle variation $(\theta_i \to \theta_o)$ and radial distances $r$ remain same. Also generalised momentum is mass times four velocity then

$$\Delta\left[p_\mu \frac{dx^\mu}{d\tau}\right] \to \Delta\left[g_{\mu\nu} \frac{dx^\mu}{d\tau} \frac{dx^\nu}{d\tau}\right] = 0 \qquad (44)$$

For diagonal metric systems Eq.(44) implies

$$\Delta\left[-g_{00}\left(\frac{dx^0}{d\tau}\right)^2\right] = \Delta\left[g_{ii}\left(\frac{dx^i}{d\tau}\right)^2\right] \qquad (45)$$

$$\Delta[-g_{00}c^2] = \Delta\left[g_{ii}(v^i)^2\right] \quad (46)$$

Using the 1st diagonalized *LT* matrix we have the variational equation

$$\Delta_\theta[(1 - b - A(r,\theta))c^2] = \Delta_\theta[(1 + b + A(r,\theta))(v^x)^2 + (1 + b)(v^y)^2 + (1 + b)(v^z)^2] \quad (47)$$

Considering others two simultaneous solutions for perturbed y and z components respectively

$$\Delta_\theta[(1 - b - A(r,\theta))c^2] = \Delta_\theta[(1 + b)(v^x)^2 + (1 + b + A(r,\theta))(v^y)^2 + (1 + b)(v^z)^2] \quad (48)$$

$$\Delta_\theta[(1 - b - A(r,\theta))c^2] = \Delta_\theta[(1 + b)(v^x)^2 + (1 + b)(v^y)^2 + (1 + b + A(r,\theta))(v^z)^2] \quad (49)$$

Adding Eqns. (47), (48) and (49) taking $v^2 = \sum_i^{x,y,z}(v^i)^2$

$$3\Delta_\theta[(1 - b - A(r,\theta))c^2] = \Delta_\theta[(1 + b + A(r,\theta))v^2] + \Delta_\theta[2(1 + b)v^2] \quad (50)$$

In Eq. (50) clearly the last square bracketed term is unperturbed thus performing variation of this part of velocity will be vanished, only the perturbed first term in connection to angular momentum $A(r,\theta)$ in right will give significant variation!

Putting for covariant velocity in quadratic $[(1 + b + A(r,\theta))(v)^2] = \tilde{V}^2(\theta)$ in Eq. (50)

$$-3[\Delta_\theta A(r,\theta)c^2] = \Delta_\theta[\tilde{V}^2(\theta)] \quad (51)$$

Dividing Eq. (51) by $\tilde{V}^2(\theta)$ we have

$$-\frac{3\Delta_\theta A(r,\theta)}{2\tilde{V}^2/c^2} = \frac{\Delta \tilde{V}}{\tilde{V}} \quad (52)$$

Now to get contravariant coordinate velocity rewriting covariant velocity in Eq.(50)

$$\tilde{V} = (1 + b + A(r,\theta))^{\frac{1}{2}} v \quad (53)$$

$$\Delta \tilde{V} = \frac{1}{2}(1 + b + A(r,\theta))^{-\frac{1}{2}}[\Delta_\theta A(r,\theta)]v \quad (54)$$

$$\text{and, } \Delta_\theta A(r,\theta) = A(\sin\theta_o - \sin\theta_i) \quad (55)$$

$$[\Delta_\theta A(r,\theta)]v = \Delta v \quad (56)$$

Putting back Eq.(56) to Eq.(54)

$$\Delta \tilde{V} = \frac{1}{2}(1 + b + A(r,\theta))^{-\frac{1}{2}} \Delta v \quad (57)$$

Using Eq.(53) and(57) together in Eq.(52) we get,

$$-\frac{6\Delta_\theta A(r,\theta)}{2v^2/c^2} = \frac{\Delta v}{v} \quad (58)$$

In regard of virial potential, $2v^2/c^2 = \frac{2GM}{c^2 r} = b$; using this relation in Eq. (58) and implying from Eq.(23) $\left(\frac{A}{b}\right)_\oplus = \frac{j}{cr}\sin\theta$, the Earth angular momentum per unit mass $j_\oplus = \frac{2}{5}\omega_\oplus R_\oplus^2$

$$\frac{\Delta v}{v} = -6\left(\frac{A}{b}\right)_\oplus (\sin\theta_o - \sin\theta_i) = 6\frac{j}{cr}(\sin\theta_i - \sin\theta_o) \quad (59)$$

Finally the flyby experiences a velocity deviation at a height of $h = (r - R)$ in replacing polar angle $\theta$ to equatorial angle $\delta$

$$\left.\frac{\Delta v}{v}\right|_\infty = 2.4\left(\frac{R_\oplus}{R_\oplus + h}\right)\frac{\omega_\oplus R_\oplus}{c}[\cos\delta_i - \cos\delta_o] \quad (60)$$

If we are little crazy further about some spacecrafts having enough anomaly like Galileo-I $(h{\sim}960km)$, NEAR$(h{\sim}539km)$ and Rosetta-I$(h{\sim}1956km)$[2, 5, 24, 25], average altitude $\langle h \rangle {\sim} 1152 km$ leads $2.4\left(\frac{R_\oplus}{R_\oplus + h}\right){\sim}2.03$ turns Eq.(60) as good as expected from Anderson's [3] empirical formula in Eq.(1) without any discrepancy just using pure General Relativistic theory.

After Multivariate fitting analysis recently L. Acedo, 2021[24] quotes Anderson's expression should include a power $\left(\frac{R_\oplus}{R_\oplus + h}\right)$ i.e. anomalies decrease with the altitude of the perigee. This is quite satisfied by the Eq.(60) surprisingly!

**Conclusions:**

This paper is purely to model the flyby anomaly in application of gravitational influence in mere General Relativistic conception. Apparently diagonalization of the Lense-Thirring weak field low angular momentum spacetime metric for rotating body has been performed. After computation the equation for osculating Earth's flyby energy anomaly shows a velocity variation Eq.(56) for in and out asymptotic velocity obeying Anderson's empirical formula! In formulating equations we are not crazy for air viscous drag not the influence of others heavenly bodies near about! The treatment going through the paper demands a sole justification of post Newtonian gravity to address the planetary anomalous energy shifting.

**Data Availability Statement:** Data sharing not applicable to this article as no datasets were generated or analysed during the current study.

**Declarations:** For this work there is no Funding and/or Conflicts of interests/Competing interests.

**References:**

[1] Antreasian, P.G. and J.R. Guinn, (1998), Paper no. 98-4287 presented at the AIAA/AAS Astrodynamics Specialist Conf. and Exhibition, Boston, 1998.

[2] Anderson, J.D., Campbell, J.K. & Nieto, M. M., (2006), NewAstron.12:383-397, 2007, http://arxiv.org/abs/astro-ph/0608087v2

[3] Anderson, J. D., Campbell, J. K., Ekelund, J. E., Ellis, J.& Jordan, J. F., (2008) Phys. Rev. Lett. 100 (2008) 091102

[4] Iorio. L., 2018, International Journal of Modern Physics D, http://arxiv.org/abs/1311.4218v2

[5] Turyshev, S. G. & Toth, V. T., (2009) http://arxiv.org/abs/0907.4184v1

[6] Antreasian P.G. & Guinn, J.R., (1998) http://www2.aiaa.org/citations/mp-search.cfm

[7] Moffat, J. W.(2006) JCAP 2006, 004 (2006) https://arxiv.org/abs/gr-qc/0506021

[8] Adler, S. L., (2008) http://arxiv.org/abs/0805.2895v4

[9] Acedio, L., 2018, http://arxiv.org/abs/1505.06884v1

[10] Goenner H. F. M., (2004) On the History of Unified Field Theories, Living Rev. Relativity 7, (2004)

[11] Hehl F. W. et al., (1976)General Relativity with Spin and Torsion: Foundations and Prospects, Rev. Mod. Phys. 48, 393-416 (1976).

[12] Poplawski N. J., (2010) Cosmology with torsion: An alternative to cosmic inflation, Phys. Lett. B 694, (3), 181-185 (2010). arXiv:1007.0587

[13] Blau, M., 2021, Lecture Notes on General Relativity, 514, 66-69, 247-249
http://www.blau.itp.unibe.ch/newlecturesGR.pdf

[14] Misner, C. W. Thorne, K.S. &Wheeler, J. A. (1972) Gravitation, W.H. Freeman &Company, 1097, 322-323,

[15] Weinberg, S., (1971), Gravitation and cosmology, John Wiley and sons, 180-181

[16] Teukolsky, S. A. (2015) *Class. Quantum Grav.* **32** 124006 **DOI** 10.1088/0264-9381/32/12/124006

[17] Visser, M.(2008) https://arxiv.org/pdf/0706.0622

[18] Lense, J. & Thirring, H.(1918) Zeit. Phys. 19 (1918), 156-163, https://www.neo-classical-physics.info/uploads/3/4/3/6/34363841/lense_thirring_-_lense-thirring_effect.pdf

[19] Qiu, W.S., Lian, D. D., Zhang, P.M., (2024), https://arxiv.org/pdf/2407.06553

**[20]** https://relativitydoctor.com/wp-content/uploads/2018/06/Supplemental-Lecture-6-Rotating-Sources-Gravito-Magnetism-and-the-Kerr-Black-Hole.pdf Eq.(23)

**[21]** Berezin, V.A., Dokuchaev, V.I.& Eroshenko, Y.N.,(2024) http://arxiv.org/abs/1704.06889v2 Vaidya metric

**[22]** Tessarotto M. & Cremaschini, C., (2021) arXiv:1601.03941v1[gr-qc]2016

**[23]** Biswas, S., (2023), Physics of the Dark Universe, https://doi.org/10.1016/j.dark.2023.101403

**[24]** Acedo, L., (2021), http://arxiv.org/abs/1701.05735v1

**[25]** P´aramos · J. & Hechenblaikner, G., (2013) Planetary and Space Science, **79-80**, https://doi.org/10.48550/arXiv.1210.7333